\documentclass[12pt,preprint]{aastex}
\usepackage{amsmath}
\usepackage{color}
\usepackage{rotating}
\usepackage[normalem]{ulem}
\usepackage{amssymb}
\usepackage{url}
\usepackage[displaymath]{lineno}
\newcommand{\Rs}{\ensuremath{\mathrm{R_S}}}

\shorttitle{CME mass evolution}
\shortauthors{Feng et al.}

\begin{document}

\title{Why does the apparent mass of a coronal mass ejection increase?}

\author{Li Feng\altaffilmark{1,2},
Yuming Wang\altaffilmark{3,*}, Fang Shen\altaffilmark{4},
Chenglong Shen\altaffilmark{3}, Bernd Inhester\altaffilmark{2},
Lei Lu\altaffilmark{1}, Weiqun Gan\altaffilmark{1}}
%\email{ymwang@ustc.edu.cn}
\altaffiltext{1}{Key Laboratory of Dark Matter and Space Astronomy, Purple Mountain Observatory,
Chinese Academy of Sciences, Nanjing, Jiangsu, China}
\altaffiltext{2}{Max-Planck-Institut f\"{u}r Sonnensystemforschung, G\"{o}ttingen, lower Saxony, Germany}
\altaffiltext{3}{CAS Key Laboratory of Geospace Environment, School of Earth and Space Sciences,
University of Science and Technology of China, Hefei 230026, China}
\altaffiltext{4}{State Key Laboratory of Space Weather, National Space Science Center, Chinese Academy of Sciences,
Beijing, China}
\altaffiltext{*}{Corresponding author, Email: ymwang@ustc.edu.cn}

\begin{abstract}
Mass is one of the most fundamental parameters characterizing the dynamics
of a coronal mass ejection (CME). It has been found that CME apparent
mass measured from the brightness enhancement in coronagraph
images shows an increasing trend during its evolution in the corona.
However, the physics behind it is not clear. Does the apparent mass gain come
from the mass outflow from the dimming regions in the low corona,
or from the pileup of the solar wind plasma around the CME when it
propagates outwards from the Sun? We analyzed the mass evolution 
of six CME events. Their mass can 
increase by a factor of 1.6 to 3.2 from 4 to 15 Rs in the field of 
view (FOV) of the coronagraph on board the \textit{Solar Terrestrial Relations 
Observatory (STEREO)}. Over the distance about 7 to 15 Rs, where the coronagraph
occulting effect can be negligible, the mass can increase by a factor of 1.3 
to 1.7. We adopted the `snow-plough' model to calculate the mass contribution 
of the piled-up solar wind in the height range from about 7 to 15 Rs. For 2/3 
of the events, the solar wind pileup
is not sufficient to explain the measured mass increase. In the 
height range from about 7 to 15~\(\Rs\), the ratio of the modelled to the measured mass
increase is roughly larger than 0.55. Although the ratios are believed to be 
overestimated, the result gives evidence that the solar wind pileup probably makes 
a non-negligible contribution to the mass increase. It is not clear yet whether
the solar wind pileup is a major contributor to the final mass derived from 
coronagraph observations. However, our study suggests that the solar wind pileup 
plays increasingly important role in the mass increase as a CME moves further
away from the Sun.

\end{abstract}


\section{Introduction}

Mass is a major parameter characterizing the dynamics of a CME. However, due to the
limitation of observations, we do not know very much about the CME mass, and its
evolution from the corona to
interplanetary space. CME masses are mostly derived from images of
white-light coronagraphs \citep[e.g.,][]{Vourlidas:etal:2000},
as a CME can cause an enhanced brightness in white-light images.
However, the mass derived from white-light coronagraph images is just an
`apparent' mass of a CME. It has two components: one is the mass of plasma
really ejected from the corona, the other is the piled-up solar wind mass
due to the compression of the solar wind plasma surrounding the CME. It is
difficult to separate the two components from imaging data. Therefore, what we often calculate
is the apparent mass. A few studies have shown that the apparent mass of a CME
increases with time \citep[e.g.,][]{Colaninno:Vourlidas:2009,Bein:etal:2013}.

The statistical work by \citet{Wang:etal:2011} has shown that the CME brightness
is positively correlated with the CME speed. A natural explanation is that a
faster CME causes higher compression of ambient solar wind and therefore a
brighter signature in coronagraph images. Numerical simulations also revealed
that the mass of the high-density shell ahead of
an erupted flux rope increases as it propagates away from the Sun but the mass of
the dark cavity, where the flux rope is located, remains almost unchanged
\citep{Lugaz:etal:2005, Riley:etal:2008}.
All these studies imply that the piled-up solar wind mass may
not be negligible, and may contribute significantly to the aforementioned
phenomenon --- apparent mass gain of a CME during its propagation.

On the other hand, there is also evidence suggesting that the CME mass gain can be
attributed to the continuous outflow from the low corona \citep[e.g.,][]
{Howard:Vourlidas:2005, Bein:etal:2013, Tian:etal:2012}.
\citet{Bein:etal:2013} investigated the CME mass evolution for a set of 25 events,
and found that the apparent mass increased by about 2\% to 6\% per solar radius
based on the COR2 observations taken by \textit{STEREO} \citep{Kaiser:etal:2008,
Howard:etal:2008}. They further found that the mass centroids 
of these CMEs moved toward the Sun relative to their self-similar expansion in 
most events. The authors suggested that the reason might be that  
the rear part of the CME with higher mass might propagate slower than the front
part of the CME. They found that the higher mass at the rear was 
supplied by the outflow from the lower corona. However, in the analysis of 
\citet{Bein:etal:2013}, the fact that materials may fall back to the Sun, 
which can also cause a backward motion of the CME centroid, was not considered.
Thus, the question how much pileup and outflow contribute to the mass increase
still remains unsolved.

In this paper, we are going to answer the question why the CME apparent mass increases
in a way different from that by \citet{Bein:etal:2013}. We try to quantitatively 
figure out at which height and how
significantly the piled-up solar wind mass contributes to the CME apparent mass.
In Section~\ref{sec_obs}, we briefly describe the selection of events, their observations,
and their 3D geometric and kinematic properties. How we derive the CME mass and
the mass evolution of selected events is presented in Section~\ref{sec_mass}.
In Section~\ref{sec_gain}, we analyze the possible source of the CME mass with the
help of a solar wind pileup model and
MHD numerical simulations. The final section consists of the discussions and conclusions.

\section{Events}\label{sec_obs}

\subsection{Event Selection}
Six CME events (see Table~\ref{tb_1}) were selected from the halo CME list
(\url{http://space.ustc.edu.cn/dreams/fhcmes/}) compiled by \citet{ShenC:etal:2013}.
The criteria to choose these events are:
(1) They appeared halo in Large Angle Spectrometric Coronagraph (LASCO) on board
\textit{Solar and Heliospheric Observatory (SOHO)} spacecraft. They were also 
observed by COR1 and COR2 coronagraphs
on board \textit{STEREO} and in most cases as almost limb events. Limb CMEs are less
occulted by the coronagraph occulter, thus are favorable for
mass estimate. (2) There was no contamination from other CMEs in the region
where the investigated CME was located.
(3) There was no contamination of very bright $\mathrm{H\alpha}$ line emission
from prominence material inside the investigated CME in the coronagraph images. Mass
estimate of such contaminated CMEs is often misleading if Thomson scattering
theory is applied \citep{Mierla:etal:2011, Carley:etal:2012}.
The six CMEs cover a broad range of leading-edge speed from 450 to 1300 km s$^{-1}$
as indicated in Table~\ref{tb_1}.

The twin spacecraft \textit{STEREO} A+B and \textit{SOHO} monitor
the solar corona from three different angles of views at about 1 AU.
COR1 and COR2 on board \textit{STEREO} observe the corona from 1.4 to 4~\(\Rs\) 
and 2.5 to 15~\(\Rs\), respectively \citep{Howard:etal:2008}. LASCO C2 and 
C3 cameras on board \textit{SOHO} have a FOV
of 2 to 6~\(\Rs\), and 3.7 to 32~\(\Rs\), respectively \citep{Brueckner:etal:1995}.

\subsection{Deprojected CME Geometry and Kinematics}

The projection of a CME onto an image may significantly distort its geometric
and kinematic parameters. For a better estimate of the mass of a CME
and its evolution, we have tried to derive the deprojected three-dimensional
(3D) geometry and kinematics. Various models/methods have been developed
to retrieve realistic parameters based on multi-point imaging data,
e.g., the forward modeling \citep[e.g.,][]{Thernisien:etal:2009, Wood:etal:2009},
triangulation method \citep[e.g.,][]{Liewer:etal:2009, Temmer:etal:2009}, polarization
ratio method \citep[e.g.,][]{Moran:etal:2010}, inverse method \citep{Antunes:etal:2009},
geometric localization \citep[e.g.,][]{deKoning:etal:2009, Byrne:etal:2010, FengL:etal:2012}.
The comparison of various models could be found in, e.g., \citet{Mierla:etal:2010},
\citet{Thernisien:etal:2011} and \citet{FengL:etal:2013a}.

Here we use the forward modeling developed by \citet{Thernisien:etal:2009} to derive
the 3D geometric and kinematic parameters of the six CME events. In this method,
the graduated cylindrical shell (GCS) model is applied, which consists
of a tubular section forming the main body of the structure attached to two
cones that correspond to the `legs' of the CME. This model has six free parameters: the
propagation longitude and latitude ($\phi$ and $\theta$), aspect ratio ($\kappa$),
tilt angle($\gamma$), half angle between two legs ($\alpha$), and heliocentric
height of the CME leading edge ($h$). From the parameters $\alpha$ and $\kappa$,
we further derived the edge-on angular width ($2\delta$) and face-on angular width ($2(\alpha+\delta)$),
where $\delta=\arcsin(\kappa)$ is the half angle of the cone, i.e., the CME leg.
Note that all the heights mentioned in this work are the radial distance
measured from the solar center. These six parameters were tuned to obtain a best match
between the modeled flux rope and the observed CME from different viewpoints.
More details of this model and its application to CMEs could be found in, e.g.,
\citet{Thernisien:etal:2009}, \citet{Thernisien:etal:2011} and \citet{Gui:etal:2011}.
For the time sequence of images of each CME, we obtained a set of parameters at
each time.
For simplicity, the free parameters are set to be constants by trial and error,
except the height $h$ which is a function of time.

\begin{figure}
\noindent\includegraphics[width=15cm]{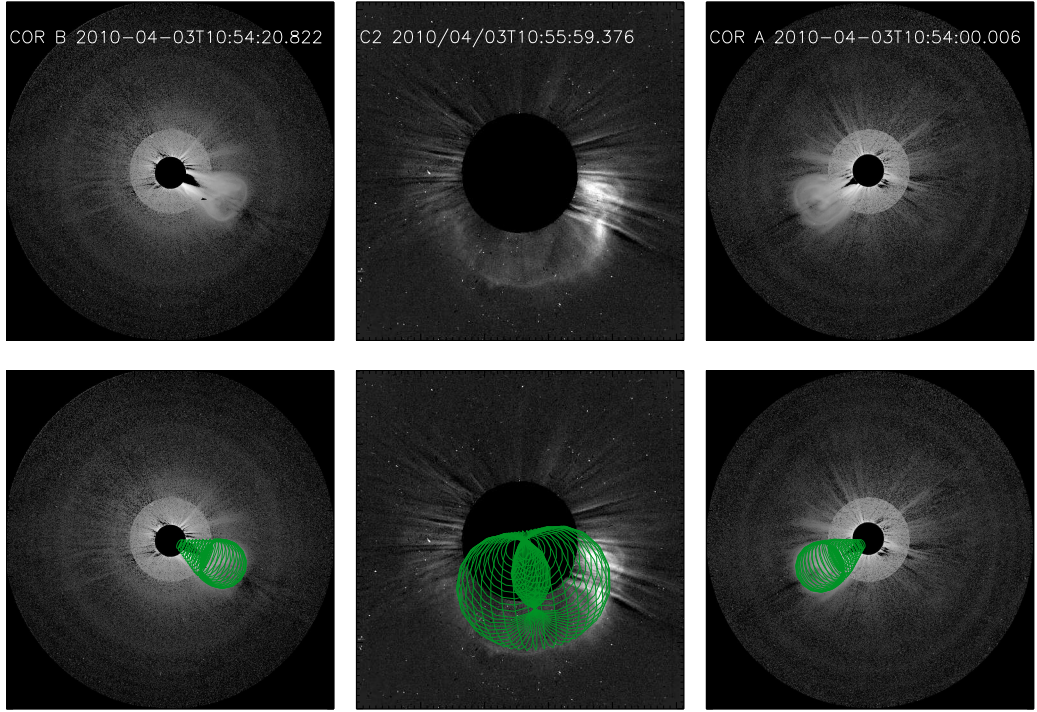}
\caption{Upper panels: the CME observed by \textit{STEREO} COR1 \& COR2 and by \textit{SOHO}/LASCO
C2 on 3 April 2010. Note that COR1 and COR2 images are incorporated to have a
comprehensive overview of the CME morphology. Lower panels: The same as the upper
panel except that the modeled flux rope indicated by green meshes are superposed.}
\label{fig:obs_gcs}
\end{figure}

\begin{figure}
\noindent\includegraphics[width=14cm]{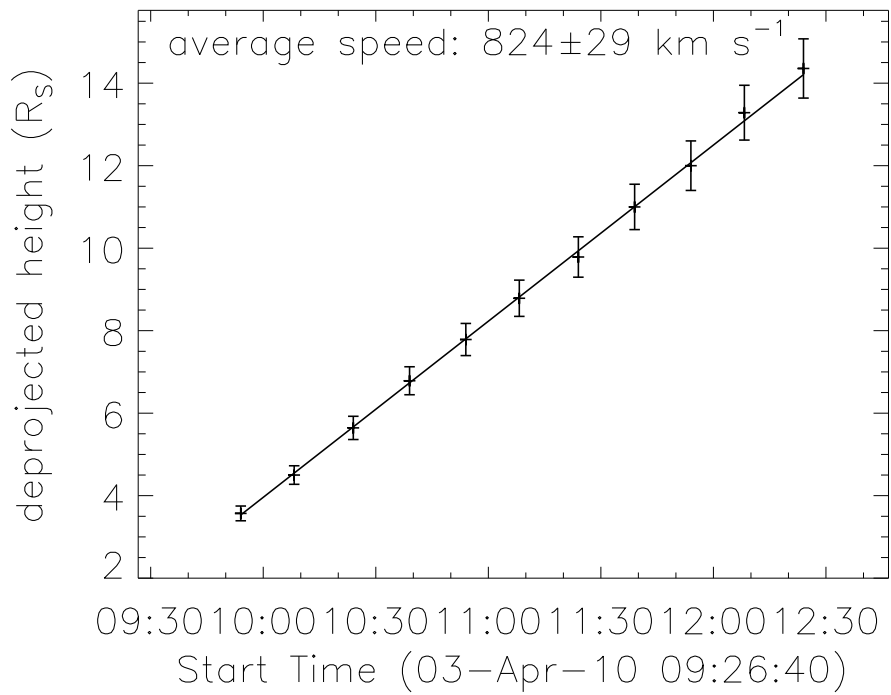}
\caption{Deprojected height as a function of time for the CME event
on 3 April 2010. The asterisks are the measurements and the black solid line is a
linear fit. The average speed derived from the linear fit is marked in the
upper part. The error bars indicate the 5\% uncertainty of the
deprojected heights.} 
\label{fig:kinematics}
\end{figure}

As a demonstration, Figure~\ref{fig:obs_gcs} shows the results for the event
on 3 April 2010. The upper panels
are the images taken by COR B, C2 and COR A, respectively, in which a pre-event image
has been subtracted to remove the background and make the CME structure more
pronounced. To get a picture of the CME as complete as possible, we incorporated
COR1 and COR2 observations. The aforementioned procedure has been applied to
all the other five CME events as well. The lower panels are the same as the
upper panels except that the green meshes are superimposed to show the best-fit
flux rope. It has to be pointed out that for the GCS 
reconstruction, we did not include in the fitting process the shock-sheath 
region when visible.
By repeatedly applying the GCS model to the time sequence of the
images of the CME, the evolution of the height of the leading edge of this CME
can be obtained and is shown in Figure~\ref{fig:kinematics}. The uncertainty in the
height is indicated by the error bars. It has two sources; one is the $\sim0.3$~\(\Rs\)
 uncertainty in the definition of the CME leading front in the white-light
images \citep[e.g.,][]{Bein:etal:2013}, and the other comes from the forward
fitting process. We assumed that the uncertainty is about
5\% of the height at each time instance. At the height of
20~\(\Rs\), it produces an uncertainty of 1~\(\Rs\)  which is close to the maximal
uncertainty in the sensitivity analysis in \citet{Thernisien:etal:2009}.
The height-time data points were further fitted by a linear function to estimate the
average speed of the CME.

Table~\ref{tb_1} lists the geometric and kinematic parameters of our six events,
i.e., longitude and latitude of the propagation direction ($\phi$ and $\theta$),
face-on and edge-on angular width ($W_f$ and $W_e$), CME solid angle ($\Omega$),
and average speed ($\mathrm{v_{avg}}$). The longitude and the latitude are in a
heliographic coordinate system with Earth sitting at zero longitude.
Positive/negative longitude represents a CME propagating westwards/eastwards.
The solid angle spanned by a CME is approximated by the multiplication of the
face-on and edge-on angular width. The leading-edge speeds of the CME events
vary from about 450 to 1300 km s$^{-1}$, showing good representativeness.

\section{Evolution of CME Apparent Mass}\label{sec_mass}

\subsection{Mass Estimate Method}
Mass calculations based on white-light coronagraph images are usually based on the
Thomson scattering theory \citep{Minnaert:1930, vandeHulst:1950, Billings:1966,
Howard:Tappin:2009}.
The major non-Thomson scattered contributions like stray light and
dust scatter are removed by subtracting a pre-event background image.
The remaining image brightness then should show the CME density
distribution produced by Thomson scattering alone. For the total brightness in
a given image pixel, the details of this scattering process can be described by

\begin{linenomath*}
\begin{gather}
b=b_\odot\frac{\pi\sigma_e}{2}
   \int_\mathrm{LOS} F(r,\chi, u) \; N_e(\mathbf{r}) \;d\ell,
\label{equ:totB}\\
    F(r,\chi,u)=2\frac{(1-u)C(r)+uD(r)}{1-u/3}
               - \frac{(1-u)A(r)+uB(r)}{1-u/3}\sin^2\chi
 \nonumber
\end{gather}
\end{linenomath*}
where
$b_\odot$ is the physical mean solar brightness (MSB) that is used as a unit
of the pixel value in calibrated COR1 and COR2 observations,
$F(r,\chi,u)$ is a function of three variables $r$, $\chi$,
and $u$. $r$ is the distance of the scattering location to the solar center,
$\chi$ is the scattering angle between the radial vector from the solar center to the
scattering electron and a position vector from the electron toward the observer.
$u$ accounts for the solar disk limb darkening, and we used the value of 0.56
for $u$ in the visible white-light spectral range.
$A$, $B$, $C$ and $D$ are known functions of $r$.
$\sigma_e$ is the differential Thomson scattering cross section, and has a value
 of $7.95\times10^{-26}\,\mathrm{cm^2\,sr^{-1}}$.

In Equation~(\ref{equ:totB}), particular attention should be paid to the
scattering angle $\chi$.
In this work, the scattering angles were computed under the assumption that a
CME is concentrated in the plane defined by the CME propagation direction and
the rotational axis of the Sun as in \citet{Colaninno:Vourlidas:2009} and \citet{
Carley:etal:2012}. And the CME propagation direction was derived in Section 2.2
using GCS forward modelling method.
Under such all-in-propagation-plane assumption of a CME, we can calculate the column
density $\int_{\mathrm{LOS}}N_e\,dl$ for any LOS with Equation~(\ref{equ:totB}).
Assuming a composition of 90\% H and 10\% He \citep{Vourlidas:etal:2000},
the mass in each pixel of a coronagraph
image can be derived by multiplying the column density with the pixel area
measured at a distance of about 1AU from the observer.

\begin{figure}
\noindent\includegraphics[width=18cm]{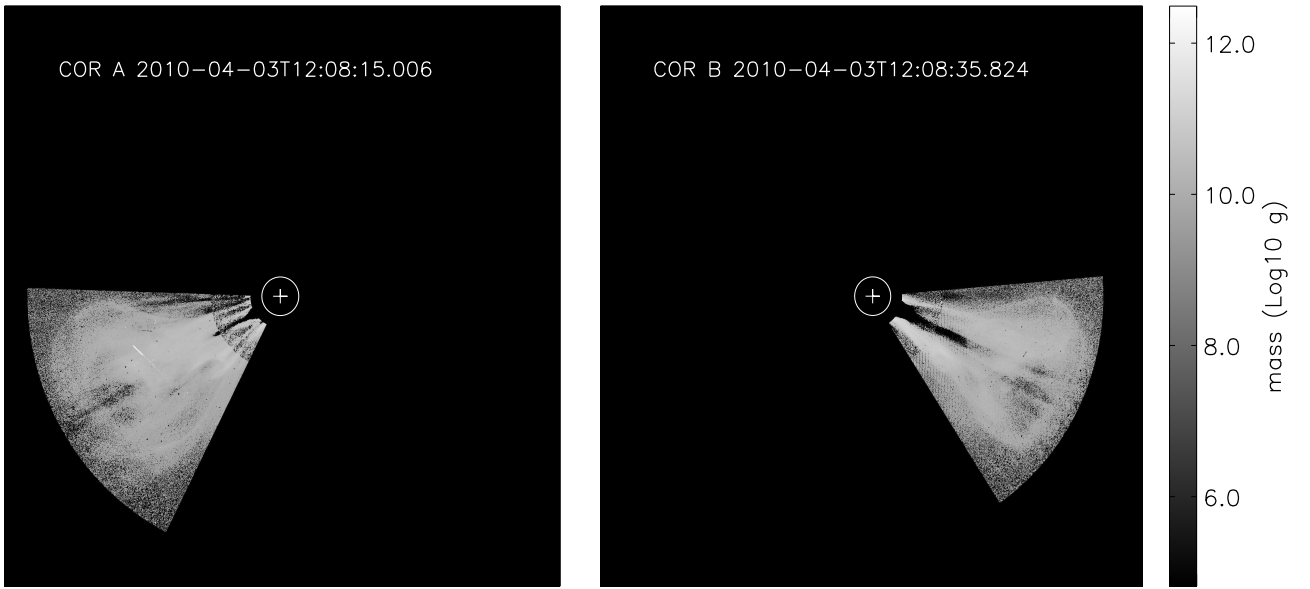}
\caption{Mass images within the CME sector regions. The plus sign and
a white circle indicate the solar center and solar disk, respectively.
The mass color bar is shown on the right side.}
\label{fig:mass_img}
\end{figure}

\begin{figure}
\noindent\includegraphics[width=16cm]{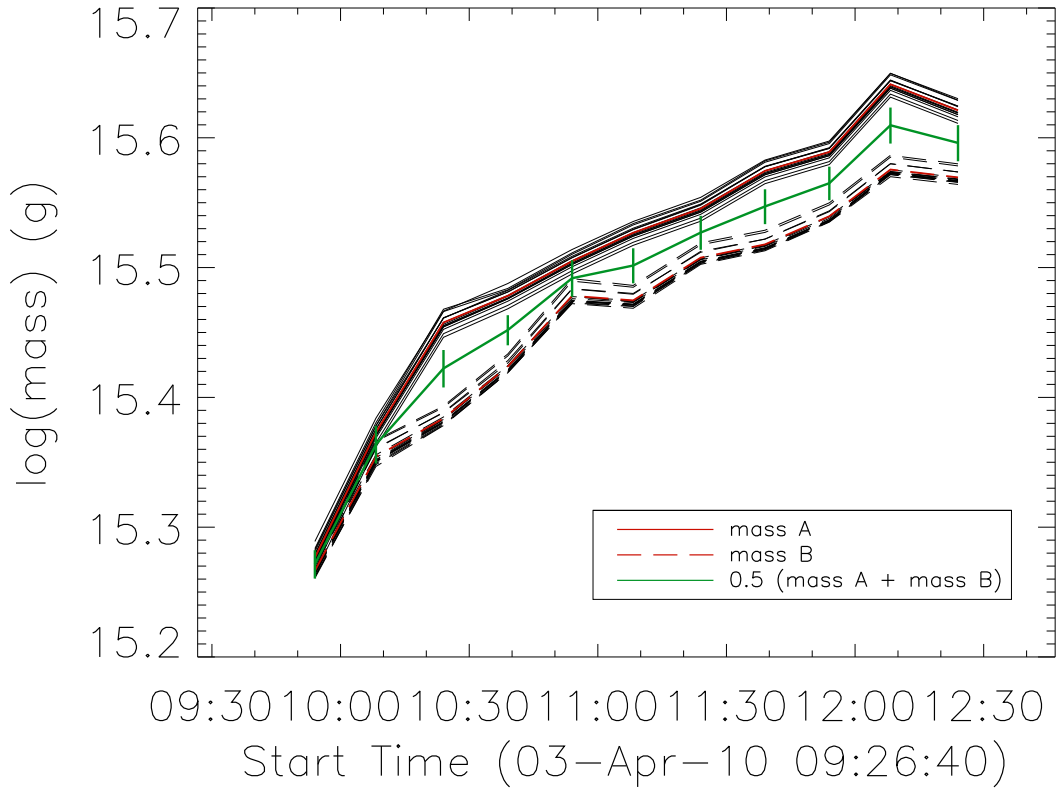}
\caption{Mass evolution with time for the CME on 3 April 2010.
The thin black solid lines are the mass derived from COR A total brightness
images with different selections of CME sectors and varing propagational
longitudes within $\phi\pm5^\circ$. A total of 15 independent mass measurements
were made for the error estimate. The mass along the thin black dashed lines
are derived in a similar way but from COR B observations. The mean mass
evolution derived from COR A and COR B are marked by solid and dashed lines in
red, respectively. 15 measurements from COR A and 15 measurements
from COR B yield 15 averaged mass over A \& B. The green line and associated
error bars are the mean and $3\sigma$ values of the 15 averaged mass measurements.}
\label{fig:mass_errAB}
\end{figure}

\begin{figure}
\noindent\includegraphics[width=15cm]{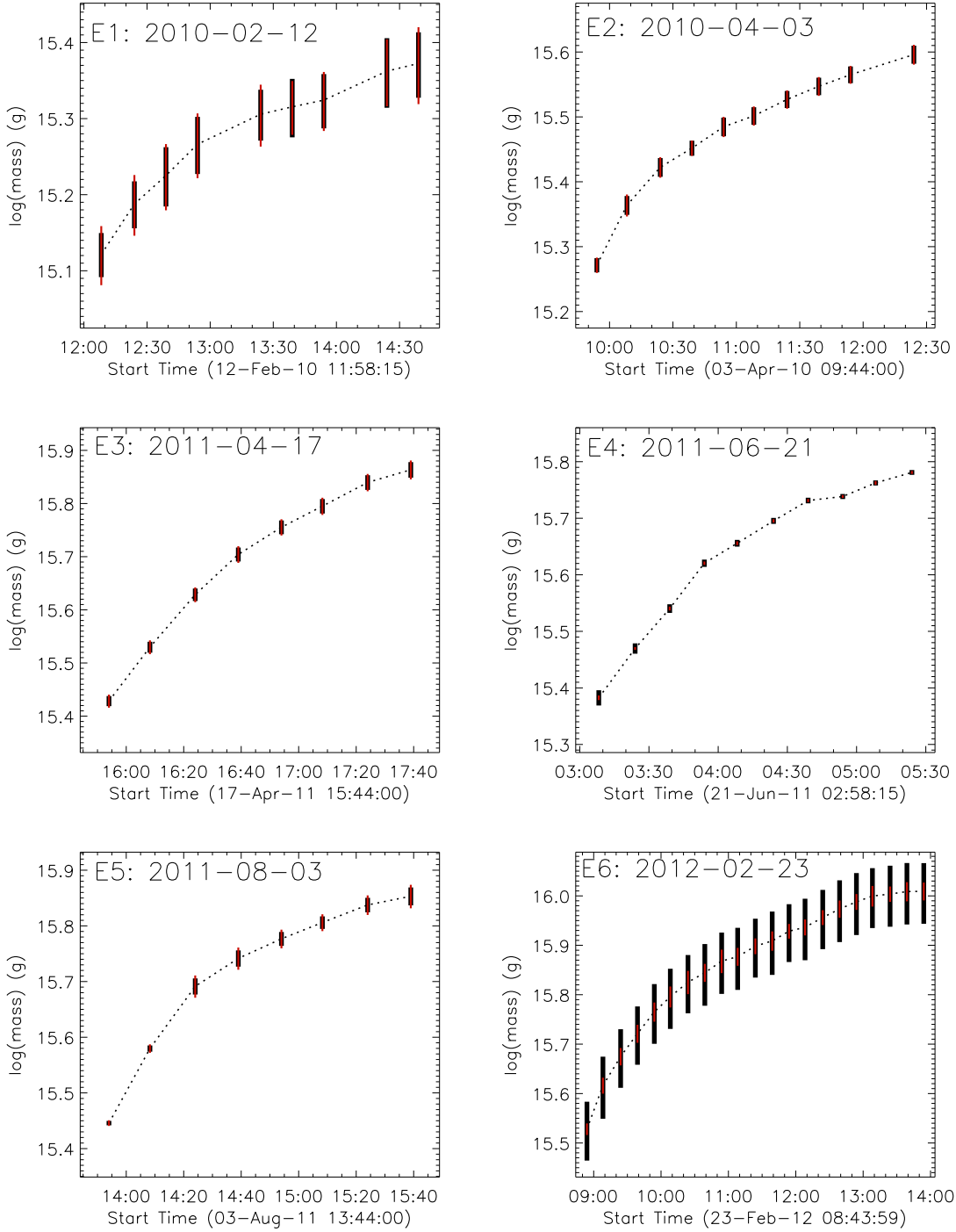}
\caption{Mass evolution of all six events. The black error bars
are the $3\sigma$ of all the 15 mass measurements with different sector regions
and different propagation angles. The red error bars are the $3\sigma$ of
eight mass measurements only with different sector regions. For most cases,
major error comes from the selection of the CME sector. While for the CME on
23 February 2012, the major error comes from the uncertainty of the propagation
angle.}
\label{fig:mass_time}
\end{figure}

In our practical procedure, we first defined a sector that contains the whole
CME. In Figure~\ref{fig:mass_img}, the mass in each pixel within the
sectors in COR A and COR B is presented for the CME occurred on 3 April 2010.
And we integrated the mass over the
entire sector and obtained two mass values from COR A and COR B data, respectively.
It should be noted that as we subtracted a pre-event background
from the image of interest, the brightness in the non-CME region is
supposed to be roughly at noise level. Thus, although the sector is larger
than the region that the CME actually occupies, the calculated values of the
apparent mass should be acceptable. Due to the different perspectives of
COR A and B, the two values are slightly different. In Figure~\ref{fig:mass_errAB}
we show the mass evolution with time. The mass evolution derived from COR A observations is
delineated by the red solid line, and from COR B observations by the red dashed
line. The average of the mass derived from COR A and B observations is
delineated by the green solid line. We found that
the difference between the mass derived from COR A and B was about $10\pm4\%$.
Some other curves and associated error bars in Figure~\ref{fig:mass_errAB}
will be explained in Section 3.2.  The evolution of the averaged mass derived from
COR A and B data for all six events is shown in Figure~\ref{fig:mass_time}.
The mass uncertainties will also be described in Section 3.2.

\subsection{Mass Uncertainty Analyses}

\citet{Vourlidas:etal:2010} has shown that the instrumentally photometric
uncertainty is quite small, and the subtraction of a pre-event image
introduces insignificant errors, if the pre-event image is selected carefully.
Similar to \citet{FengL:etal:2015}, we mainly considered two sources of error
in the mass estimate.
One is the LOS distribution of the CME mass in Equation~(\ref{equ:totB}),
especially the longitudinal propagation angle $\phi$. The other source is our
somewhat arbitrary selection of the CME sector. To estimate the uncertainties
from these two sources, we varied $\phi$ within $\phi\pm5^\circ$ at seven values
and repeated the CME sector selection process independently for eight times.
A total of 15 measurements were made from COR A and B observations, respectively.
They are shown by black solid and black dashed lines in Figure~\ref{fig:mass_errAB}.
Correspondingly, the red solid and dashed lines are the averaged values over 15
measurements. For our analyses of the mass evolution of a CME event, we adopted
the green curve. 15 measurements from COR A and 15 measurements
from COR B yield 15 averaged mass over A \& B. The green line and associated
error bars are the mean and $3\sigma$ values of the 15 averaged mass measurements.

Figure~\ref{fig:mass_errAB} presents an example of the mass evolution and
error analyses. In Figure~\ref{fig:mass_time}, the mass evolution and
associated error bars of all six events are shown. The black error bars are
the $3\sigma$ of all 15 mass measurements with different sectors and
propagation longitudes, while the red error bars are the $3\sigma$ of the mass
measurements only with different sector regions. In most cases the major uncertainty
comes from the selection of the CME sector. However, for the CME observed on
23 February 2012, the major uncertainty originates from the error of the
propagation longitude.

\begin{figure}
\noindent\includegraphics[width=15cm]{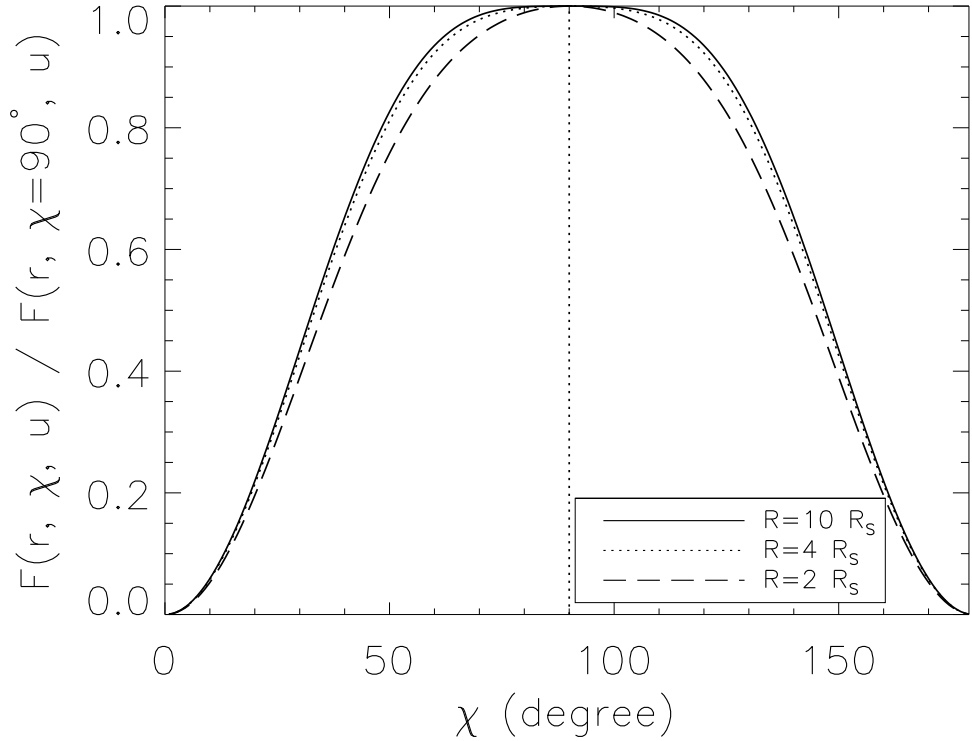}
\caption{Dependence of the function $F(r, \chi, u)$ on the
scattering angle $\chi$ at three impact distances (R=2, 4, 10 \(\Rs\)).
For COR observations, the impact distance is the projected
distance of $r$ from the scattering location to the solar
center onto the POS. The brightness is normalized by its maximal value at the POS
($F(r, \chi=90^\circ, u)$). The vertical dotted line marks the position of
$\chi=90^\circ$.}
\label{fig:norm_B}
\end{figure}

To explain the results of uncertainty above, we include in Figure~\ref{fig:norm_B}
the dependence of the function $F(r,\chi,u)$ on the scattering angle $\chi$ at
three impact distances of 2, 4, and 10 \(\Rs\). The ordinate is $F(r,\chi,u)$
normalized by its maximum at the POS $\chi=90^\circ$. For COR observations, the impact
distance is the projected distance of $r$ from the scattering location to the
solar center onto the POS. As revealed by Figure~\ref{fig:norm_B}, when $\chi$
deviates from $90^\circ$ within about $30^\circ$, $F(r,\chi,u)$ only changes
slightly and forms a plateau around $\chi=90^\circ$. This phenomenon has also
been found by \citet{Howard:DeForest:2012} where they termed it ``Thomson plateau''
for the heliospheric HI observations and 
\citet{Colaninno:Vourlidas:2009}. Therefore, a small change in propagation
longitude $\phi\pm5^\circ$, which causes a small change of $\chi\pm5^\circ$,
does not effect $F(r,\chi,u)$ much within the $30^\circ$-deviation range.
Subsequently, the column density and mass for a given pixel derived from
Equation~(\ref{equ:totB}) does not change much when $\chi$ deviates from $90^\circ$ within
about $30^\circ$. This is the case for the first five CME events. Therefore, the
uncertainty of the mass for these events mainly comes from the selection of
the CME sector rather than the propagation longitude.

In the case of the CME occurred on 23 February 2012, the propagation direction
deviated from the POS as seen by \textit{STEREO} A by about $50^\circ$, and from the POS
as seen by \textit{STEREO} B by about $90^\circ$. As the CME appeared in the FOV of COR B
as a full halo, the corresponding mass calculation subjects to a large error.
Therefore, we only used COR A data for mass analyses. When we derived the
mass from COR A data, a small change in $\chi$ yielded a large change in mass.
It is because the slope of $F(r,\chi,u)$
at $\chi\approx140^\circ$ is quite steep in Figure~\ref{fig:norm_B}, a small
change of $\chi\pm5^\circ$ makes a large change in $F(r,\chi,u)$, hence in column
density and mass. That is why the major uncertainty of this event comes from
the uncertainty of the propagation longitude.

Note that our all-in-propagation-plane assumption yields an infinitesimal angular width
for a CME. Mass can be underestimated under such an assumption according to
\citet{Vourlidas:etal:2000}. The authors derived the ratio of the mass estimate
under the infinitesimal-angular-width assumption to the mass under a finite-angular-width
assumption as a function of the angular width of a CME. As the angular width
usually keeps nearly constant in time in the COR FOV, the under-estimate ratio
should also keep constant. Therefore, the mass change rate $dm/dt$ is not
effected by the infinitesimal-angular-width assumption.

\subsection{Treatment of the Occulting Effect}
Figure~\ref{fig:mass_time} clearly shows that the mass evolution has an
increasing trend. \citet{Bein:etal:2013} have demonstrated that one of the major
causes of such an increase is the entering of the CME ejecta into the COR FOV
from behind the occulter, which
we call occulting effect. They then showed that the apparent mass evolution
could be fitted by the following function
\begin{linenomath*}
\begin{equation}
m(h)=m_0\left[1-\left(\frac{h_{occ}}{h}\right)^3\right]+\Delta m(h-h_{occ}),
\label{equ:m_h}
\end{equation}
\end{linenomath*}
where the parameter $h_{occ}$ is the height of the outer edge of an effective occulter,
$m_0$ is a modeled initial apparent mass of a CME, and $\Delta m$ is an
additional mass increment per unit height. The first term on the right-hand side
of Equation~(\ref{equ:m_h})
represents the increase of visible mass from a homogeneous cone-shaped and
self-similarly expanding CME. It describes how the visible mass of a CME with a
constant mass $m_0$ evolves, when entering the FOV of COR due to its expansion.
The increase of the visible mass above the occulter is merely a geometric effect.
The second term is an empirical correction to this simplified model evolution,
as the CME mass could increase due to physical reasons, e.g., mass outflow from
a dimming region, or solar wind pileup. In the model of
\citet{Bein:etal:2013}, the actual physical
mass increase of a CME can be approximated by $m(h)=m_0+\Delta m(h-h_{occ})$.

\begin{figure}
\noindent\includegraphics[width=14cm]{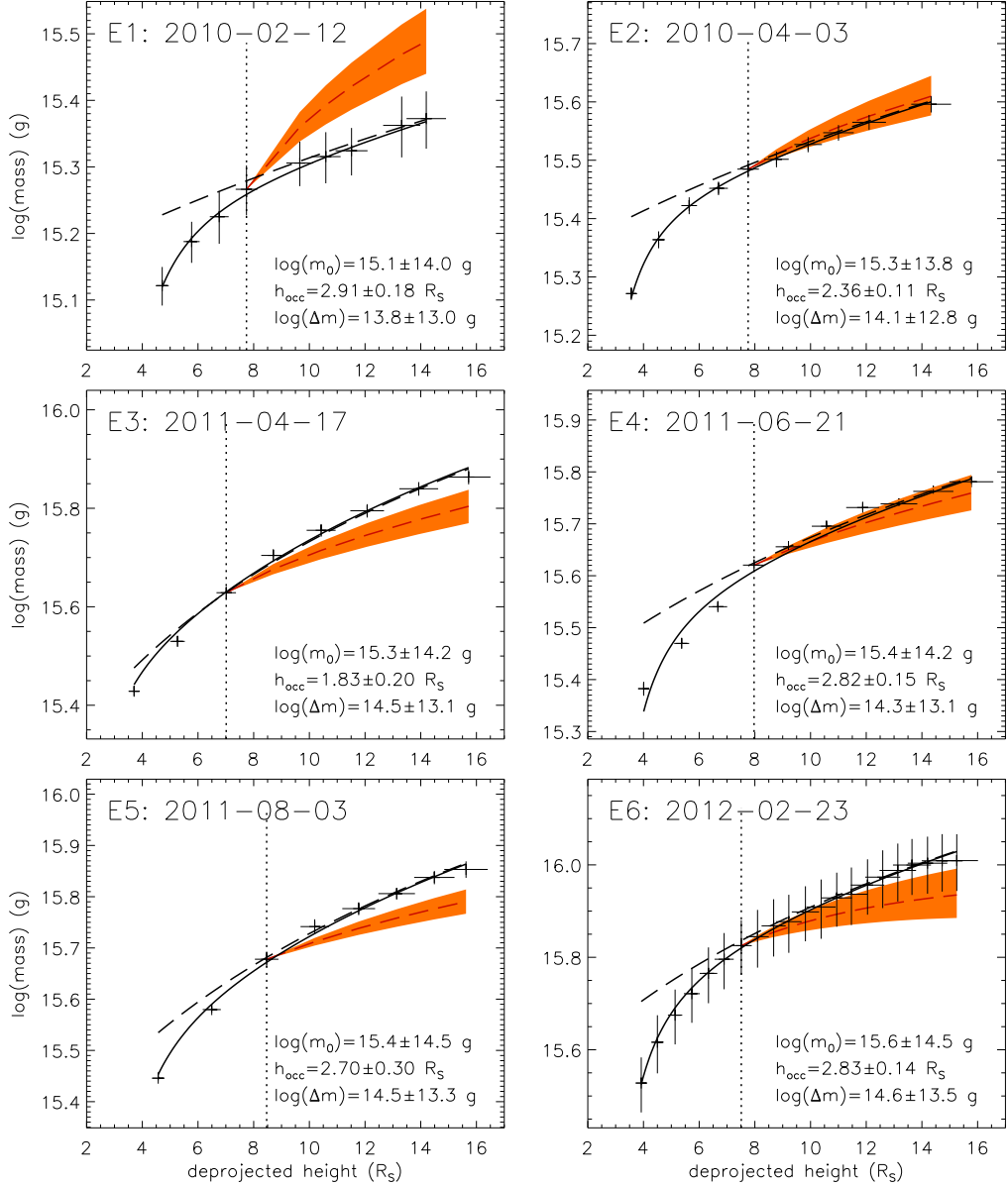}
\caption{Mass profiles as a function of the deprojected height.
The best fit result of Equation~(\ref{equ:m_h}) is indicated by a black solid line in each
panel. The mass evolution after removing the occulting effect is shown by
the black dashed line. The best fit parameters are marked in the
bottom right in each panel. For $m_0$ and $\Delta m$, the mean$\pm\sigma$ is
written in logarithmic scale. The vertical dotted line marks the height $h_1$
where the occulting effect starts to be negligible. The mass
evolution due to the solar wind pileup above $h_1$ is shown by a red dashed
line in each panel. The uncertainties indicated by the shadow region in orange are 
derived under the assumption that $n_{sw}$ and $v_{sw}$ vary in the range from 80\% 
to 120\% of the simulated values.}
\label{fig:mass_heig}
\end{figure}

In Figure~\ref{fig:mass_heig}, the black discrete data points are the calculated mass
as a function of deprojected height with the error bars in horizontal and
vertical directions. Taking these error bars into
account, we fitted Equation~(\ref{equ:m_h}) to the mass-height measurements
of each CME event. The best-fit parameters for each CME event are written in
the bottom right, and for $m_0$ and $\Delta m$, the mean $\pm \sigma$
are in logarithmic scale.
In each panel the fitting results are shown by the black solid
line, and the black dashed line is the mass variation $m_0+\Delta m(h-h_{occ})$
with the occulting effect removed. The vertical line marks the height from which
the occulting effect starts to be negligible. It is defined as the height from
which the measured mass can reach above 97\% of the dashed line. It is found that
the occulting effect could be ignored once the CME reached about 7 to 8 \(\Rs\).
We denote this height as $h_1$, the corresponding mass at $h_1$ is $m_1$. $m_1$
and $h_1$ for all six CME events are listed in Table~\ref{tb_2}.
The other two mass quantities that are also included in Table~\ref{tb_2} are
$m_0$ below $h_{occ}$, and the final apparent mass $m_2$ before a CME leaves
COR FOV at $h_2$.

Note that our purpose of adopting this model is only to obtain the height where the
occulting effect starts to be negligible. We did not use this simplified model
for further analyses of mass evolution. The measured mass evolution may
deviate from the linear increase assumption in Equation~(\ref{equ:m_h}). Actually,
in Figure~\ref{fig:mass_heig}, for most events, when the CME starts to leave 
the COR FOV, the measured mass is lower than this linear-model prediction.

\section{Sources of CME Apparent Mass Gain}\label{sec_gain}

As mentioned before, the CME apparent mass might consist a significant
contribution from ambient compressed solar wind plasma. In some appropriate
circumstances, shock might form around a CME moving faster than the fast
magnetoacoustic wave. For events E3, E5, in COR images we saw deflected
streamers which were not adjacent to the CME, and left a very diffusive space
between the deflected streamer and the CME. It might be the signature of a
shock \citep{Vourlidas:etal:2003, Ontiveros:Vourlidas:2009}.
It is almost impossible to separate the brightness contribution of the shock
from that of the CME. Although solar wind pileup and compression due to a MHD
shock are different physical processes, as the observed shocks were far fainter
than the CME, we did not remove these possible shocks from our mass analyses and
treated them as an extreme case of compression.

To quantify the piled-up mass of the solar wind plasma surrounding the CME, a
highly ideal model, called the `snow plough' model, is applied. In this model,
a CME accretes mass as it propagates and its momentum is changed through the
interaction with the ambient solar wind. The mass change
per unit time is described by (see also in \citealt{Tappin:2006})
\begin{eqnarray}
\frac{dm}{dt} = \rho_{sw}A|v_{sw}-v|\label{eq_pileup2},
\end{eqnarray}
where $m$, $v$ and $A$ are the mass, speed and the cross-sectional area
of a CME, respectively, and $v_{sw}$ and $\rho_{sw}$ are the ambient solar wind
speed and density, respectively.
The value of $A$ is approximated by the solid angle $\Omega$ in Table~\ref{tb_1}
times $h^2$, i.e.,$A=4\delta(\alpha+\delta)h^2$. The derivation of the above
equation could be found in Appendix~\ref{app_1}.
It should be noted that the equation assumes that the entire piled-up plasma
moves together with the CME.
Although this assumption may not be true, it is still worth investigating
if this model can represent the mass gain during the CME propagation.
Actually the `snow plough' equation gives the maximal amount of mass that the
solar wind can contribute to CME mass.

\subsection{Parameters of Background Solar Wind}

The `snow plough' model requires the solar wind parameters. Thus we use 3D MHD
simulations to obtain the undisturbed solar wind parameters along the CME path
from 2.5 to 20~\(\Rs\). The solar wind density and speed at
different heights are calculated as the average over a spherical sector of 60 degrees
which is centered at the 3D propagation direction of a CME. The numerical
scheme in the 3D MHD simulations is a total variation diminishing (TVD) scheme
in a Sun-centered spherical system \citep{FengX:etal:2003,FengX:etal:2005,
ShenF:etal:2007, ShenF:etal:2009}. For all six
events, the synoptic maps of longitudinal magnetic field over a Carrington rotation
from Wilcox Solar Observatory are used as input to the code for reality.
The details of how to get a steady background solar wind could be found in our
previous works \citep[e.g.,][]{ShenF:etal:2013, Wang:etal:2014}.

\begin{figure}
\noindent\includegraphics[width=15cm]{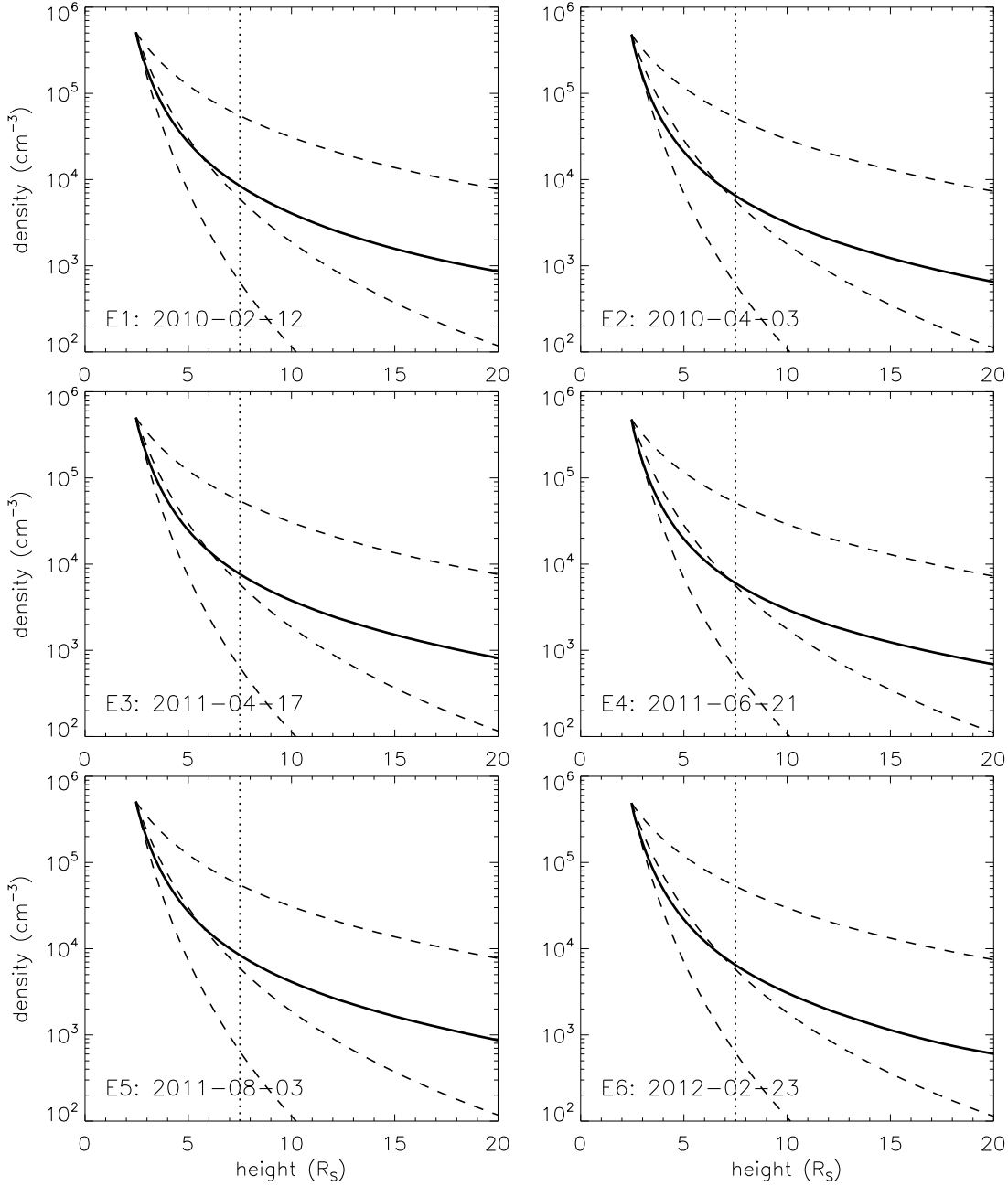}
\caption{The solar wind number density against height derived from MHD
simulations. The number density is the average of the density over a cone of 60 degrees
centered at the CME propagation direction. Three dashed lines indicate the variation of
density following $h^{-2}$, $h^{-4}$, and $h^{-6}$, respectively, from upper
to bottom. The vertical dotted lines mark the height at 7.5~\(\Rs\).}
\label{fig:sw_den}
\end{figure}

\begin{figure}
\noindent\includegraphics[width=16cm]{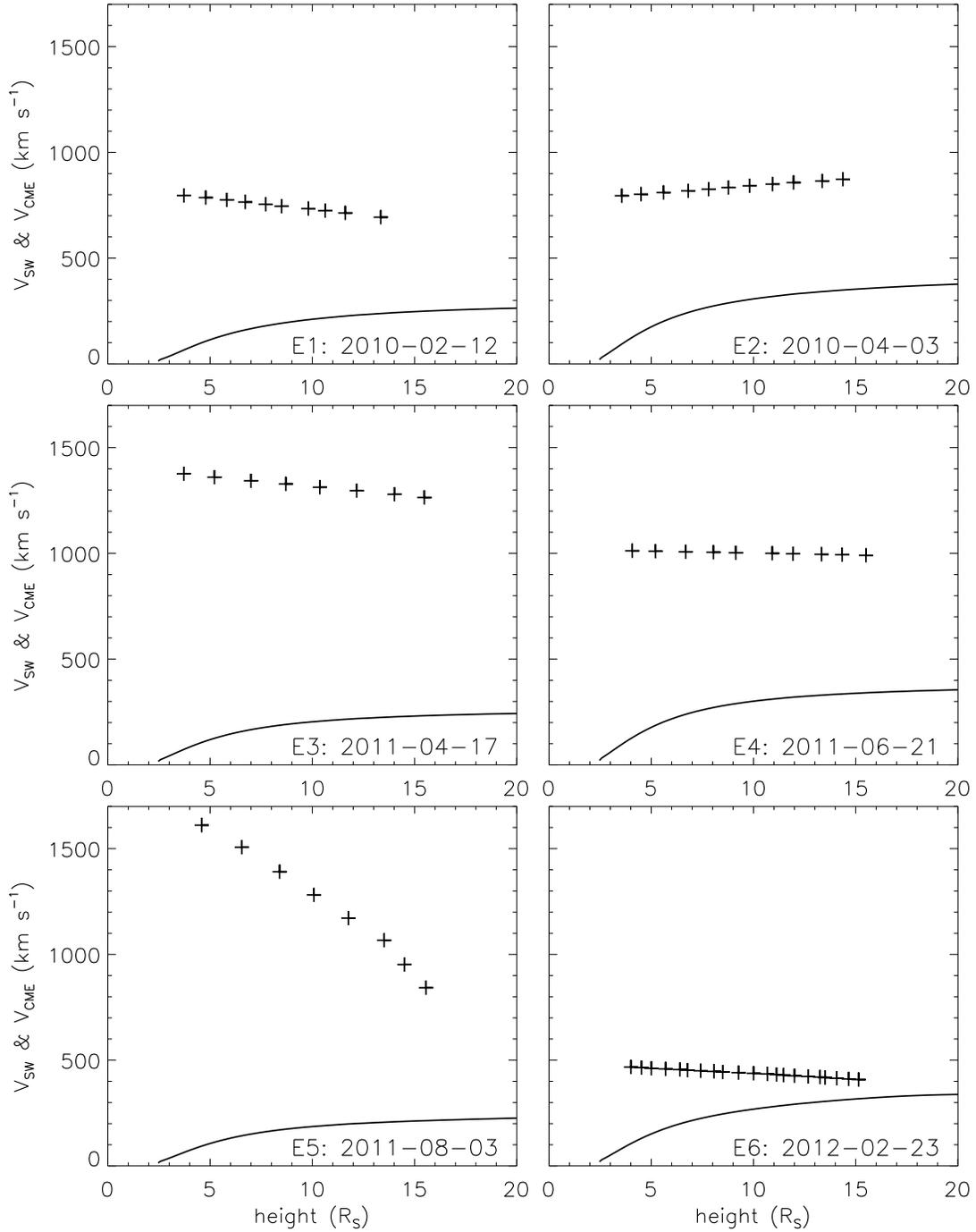}
\caption{Solid lines: the solar wind radial speed against height derived from MHD
simulations, which is the average of the speed over a cone of 60 degrees 
centered at the CME propagation direction. Plus signs: the CME speed.}
\label{fig:sw_vr}
\end{figure}

The averaged solar
wind density and speed against height over the spherical sectors are presented
in Figure~\ref{fig:sw_den} and \ref{fig:sw_vr}, respectively. For all six events,
the variation of density follows about $h^{-6}$ to $h^{-4}$ below 7 to 8~\(\Rs\),
and follows $h^{-2}$ above 7 to 8~\(\Rs\). It is also the height where the occulter
effect starts to be negligible. The solar wind speeds have a rapid increase at
lower heights and then a slower increase at larger heights. The speeds have a similar trend of
evolution for different Carrington rotations, but reach different values at
20~\(\Rs\) which range from about 220 to 460 km s$^{-1}$. The solar wind parameters
at 20~\(\Rs\) have been listed in Table~\ref{tb_2} for reference. 
We also plot in each panel the speed 
of the CME derived from a quadratic fit to the height-time diagram. The scale of 
the solar wind speed is marked on the left $Y$ axis, and the scale of the
CME speed is marked on the right $Y$ axis. For all events, we find that 
$v_{\mathrm{cme}}-v_{\mathrm{sw}}$ decreases with height.

\subsection{How Much Mass Can Solar Wind Pileup Contribute?}

The calculation of the mass that the solar wind pileup can contribute is
straightforward, as all the parameters $m$, $v$, $A$, $\rho_{\mathrm{sw}}$ and 
$v_{\mathrm{sw}}$
can be obtained from either CME observations or MHD numerical simulations.
The CME speed $v$ was estimated from a quadratic fit to the height-time 
plot, as it provides a better result than the linear fit.
We started the mass pileup from $m_1$ at $h_1$, below which we assumed that
the mass increase is due to the geometric occulting effect and the mass
supply from lower corona. If the mass increase above $h_1$ was solely
because of the solar wind pileup, we would obtain the mass evolution indicated
by the red line in each panel of Figure~\ref{fig:mass_heig}. The shadow region in 
orange is obtained by assuming that the solar wind density and velocity have an 
uncertainty of 20\%.

It is found that, for most events (E3 through E6), the solar wind pileup mass is 
not sufficient to interpret the measured mass increase from COR observations 
represented by black symbols, suggesting that apparent mass increase is also 
contributed by the outflow from the low corona. For the event E1, the solar wind
pileup mass has a higher value than the measured ones above $h_1$ in the late-phase 
evolution in the COR FOV, while for the event E2, it is more or less consistent 
with the measured mass. Note that the `snow plough' model gives an upper limit of 
the mass estimate. This model assumes that the solar wind 
plasma colliding with the CME will all be attached to the CME. However, this is
an ideal case. Some solar wind plasma may not be attached, and just flow around
the CME and eventually become the ambient background. Therefore, the 
actual mass increase due to the solar wind pileup may drop below the red lines.

Although the solar wind pileup mass may be overestimated, the results still implies
that the solar wind pileup probably makes a significant contribution to the apparent
mass increase observed by the coronagraphs. By comparing the piled-up 
mass $m_s$ from $h_1$ to $h_2$ with the total mass gain $m_2-m_1$ during 
the same period, we find that
the piled-up mass occupies more than half of the total mass gain as listed in
the column for $f_{s}$ in Table~\ref{tb_2}, or $m_s$ is at least 19\% of the 
final mass $m_2$ before 
the CMEs leave the COR FOV. Since the piled-up mass are overestimated
and our infinitesimal-angular-width assumption may make the measured mass underestimated,
the ratios of the piled-up mass to the total mass gain given above are obviously overestimated.
However, even if we consider a 100\% overestimate, the ratios of the piled-up mass is
larger than 25\% in the height range from $h_1$ to $h_2$. That is to say, although
the pileup is not sufficient to explain the observed mass increase, its contribution
is non-negligible. Due to the overestimation, it is not clear whether the pileup
could make a major contribution to the measured mass increase.

Besides, from Figure~\ref{fig:mass_heig}, we can find that the slopes of mass 
evolution $dm/dh$ of the
measured mass and modeled pileup mass gradually decrease with height. When the CME was
about to leave the COR FOV, the two slopes are
nearly the same for most of the events except the first one. It indicates that
solar wind pileup makes a more important contribution to
the mass increase at larger distances from the Sun.

\section{Conclusions and Discussions}
Based on the brightness enhancement in coronagraph images, we have followed
the mass evolution of six CMEs. The deprojected speed of
the investigated CMEs covers from 450 to 1300 km s$^{-1}$. All these CMEs with
different kinematics have an increasing trend in their mass evolution even when
the occulting effect is removed. The long-lasting accumulation 
of CME mass in coronagraph FOV implies that the initial kinetic energy in the 
CME source region is smaller than the energy estimate using the final asymptotic mass
\citep{Carley:etal:2012, FengL:etal:2013b}.

Physically, there may be two sources
of CME mass gain in the corona: the solar wind mass piled up around the CMEs
and the mass supply by the outflow from the dimming regions in the low corona
\citep{Jin:etal:2009, Tian:etal:2012, Aschwanden:etal:2009}. 
We calculated the mass contribution of the solar wind pileup from the height
beyond which the occulting effect is negligible. It is found that solar wind pileup
probably may make a non-negligible contribution to the apparent 
mass increase observed
by coronagraphs. For all of the events, the solar wind piled-up mass might occupy
more than half of the total mass gain during the same period from 7 or 8 \(\Rs\) to the
edge of the FOV of COR2. It has to be pointed out that those 
ratios represent upper limits of the pileup contribution. However, 
even if we consider a 100\% overestimation, the ratio
of the pileup mass to the total mass gain could be larger than one fourth in 
the height range from 7 to 15 \(\Rs\). We also find that the
contribution from the solar wind pileup becomes increasingly significant as
a CME propagates to larger distances from the Sun. In short, Our work reveals 
that the solar wind pileup is
not sufficient to explain the measured mass increase in the COR FOV. However, 
its contribution is non-negligible. Due to the overestimation, it is not clear 
yet whether the pileup
could make a major contribution to the measured mass increase.

Whether the solar wind pileup is a major contributor to the 
final apparent mass in the COR FOV is not clear either. If we assume that the 
mass below 7 or 8 \(\Rs\) came from the mass outflow from the dimming region, 
the ratio of $m_s/m_2$ reveals that the solar wind pileup comprises less than 
half of the final mass for all events. In this case, it may not be a 
dominant contributor to the CME apparent mass in the COR FOV. Whether this 
assumption is valid or not requires further investigation. Future work involving 
the mass estimate from the dimming regions using SDO/AIA multi-wavelength data
will be pursued. If the solar wind pileup could also contribute to the mass 
below about 
7 \(\Rs\), for some of the CMEs, the pileup may be a dominant contributor.

Both the piled-up mass derived from the `snow-plough'
model and the `virtual mass' introduced in \citet{Cargill:2004} are related to 
the solar wind mass density. However, they are theoretically and observationally different.
From hydrodynamic point of view, the `virtual mass' is the `added mass' of an 
accelerating body moving in a fluid, and is required to obtain the correct 
accelerating force in the momentum equation. The `virtual mass' in 
\citet{Cargill:2004} was derived under the assumption of an incompressible fluid, 
which means that the density in the surrounding solar wind does 
not change over time. The total brightness observed with a coronagraph is 
proportional to the density, as there is no density change caused by the added 
mass, we will not be able to detect it from base-difference images. On the
other hand, the piled-up mass in the `snow-plough' model is a compressional 
effect, and the resulting density enhancement could in principle be 
detected in the base-difference images. Furthermore, pileup occurs even 
when a CME is moving steadily without any acceleration.

\citet{DeForest:etal:2013} tracked the CME mass from \textit{STEREO}/COR2 FOV to HI2
FOV, and found that the mass of the CME under investigation continuously increased
until to the distance of about 0.7~AU, but the mass increase dropped with
distance. It is consistent with the trend of evolution predicted from our mass
evolution in the COR FOV. When the CME started to leave the COR FOV, the measured
mass increase per height ($dm/dh$) seems to agree with the prediction of the
`snow plough' model. As an outlook of the work in this paper, we will test
whether this model can explain the mass evolution in the HI FOV.
Furthermore, if we assume a proper shape of CME at 1 AU, the mass residing in the
CME and in the shock sheath can be approximately calculated from in-situ measurements, which
can add another data point in the mass evolution profile.

\acknowledgements
\textit{SOHO} and \textit{STEREO} are projects of international cooperation
between ESA and NASA. The SECCHI data are produced by
an international consortium of NRL, LMSAL, and NASA GSFC
(USA), RAL, and U. Birmingham (UK), MPS (Germany), CSL
(Belgium), IOTA, and IAS (France). SDO is a mission of NASA's
Living With a Star Program. L.F. and W.Q.G. are supported by
the grants from MOSTC (2011CB811402), NSFC (11473070, 11427803,
11233008 and 11273065), NSF of Jiangsu Province (BK2012889), CAS(XDA04076101). 
Y.W. and C.S. are supported by the grants from MOSTC (2011CB811403),
CAS (KZZD-EW-01-4) and NSFC (41131065, 41121003 and 41274173).
F.S. is supported by the grants from MOSTC (2012CB825601) and NSFC 
(41174150 and 41474152).
L.F. also acknowledges the Youth Innovation Promotion Association, CAS, 
for financial support. 
The work of B.I. is supported by DLR contract 50 OC 1301.

\appendix
\section{`Snow plough' model for CME propagation}\label{app_1}

`Snow plough' model assumes that all the solar wind plasma colliding with the CME
will be attached to the CME. Based on this assumption and considering the 1D problem, 
the mass exchange between the CME and the solar wind is given by
\begin{eqnarray}
\Delta m=-\Delta m_{sw}=\rho_{sw}A|v_{sw}-v|\Delta t
\end{eqnarray}
or
\begin{eqnarray}
\frac{dm}{dt}=-\frac{dm_{sw}}{dt}=\rho_{sw}A|v_{sw}-v|\label{eq_1}
\end{eqnarray}
where $m$, $v$ and $A$ are the mass, speed and the area of the cross-section of the CME, respectively,
and $m_{sw}$, $\rho_{sw}$ and $v_{sw}$ are the mass, density and speed of the solar wind,
respectively. The cross-section area $A=4\delta(\alpha+\delta)h^2$, where $2\delta$
is the flux rope edge-on angular width, $2(\alpha+\delta)$ is the face-on angular width, 
and $h$ is the height of the leading edge. Using absolute value of $(v_{sw}-v)$ 
means the solar wind pileup could occur at either the frontside or backside of 
the CME. When the CME speed is higher than the solar wind speed, 
the solar wind piles up mainly at the frontside of the CME; When 
the CME speed is lower than the solar wind speed, the pileup mainly occurs at the 
backside. For the CME events in this paper, we have CME speed always higher
than the solar wind speed. Therefore the pileup mainly occurred at the frontside.

The mass exchange causes the loss of the momentum of the background solar wind, which is
\begin{eqnarray}
\Delta p_{sw}=\Delta m_{sw}v_{sw}=-\rho_{sw}A|v_{sw}-v|v_{sw}\Delta t \label{eq_2}
\end{eqnarray}
or 
\begin{eqnarray}
\frac{dp_{sw}}{dt}=\frac{dm_{sw}}{dt}v_{sw}=-\rho_{sw}A|v_{sw}-v|v_{sw}\label{eq_3}
\end{eqnarray}

Meanwhile, the conservation of momentum requires $\frac{dp}{dt}=-\frac{dp_{sw}}{dt}$, i.e.,
\begin{eqnarray}
&&\frac{dmv}{dt}=-\frac{dm_{sw}}{dt}v_{sw}\\
\Rightarrow && \frac{dm}{dt}v+ma=-\frac{dm_{sw}}{dt}v_{sw} \label{eq_4}
\end{eqnarray}
where $a$ is the acceleration of the CME. Using Eq.\ref{eq_1},
the above equation leads to
\begin{eqnarray}
ma=\frac{dm}{dt}(v_{sw}-v)\label{eq_5}
\end{eqnarray}
On the other hand, combine Eq.\ref{eq_3} and \ref{eq_4}, we have
\begin{eqnarray}
\frac{dm}{dt}v+ma=\rho_{sw}Av_{sw}|v_{sw}-v|\label{eq_6}
\end{eqnarray}
After Substituting Eq.\ref{eq_5} for $ma$ in the above equation, it is obtained that
\begin{eqnarray}
\frac{dm}{dt}=\rho_{sw}A|v_{sw}-v|\label{eq_7}
\end{eqnarray}
which is the equation we used to calculate the mass contribution from solar wind
pileup.

%\bibliographystyle{apj}
%\bibliography{cme}

\begin{table}
\caption{The geometric and kinematic parameters of six CME events.}
\centering
\begin{tabular}{cccccccc}
\hline
No. & Date & $\phi$($^\circ$) & $\theta$ ($^\circ$) & W$_{f}$ ($^\circ$) & W$_{e}$ ($^\circ$) &
$\Omega$ (sr) & $v_{\mathrm{avg}}$ (km s$^{-1}$) \\
\hline
E1 & 2010-02-12 & -2.5   & 7.2  & 68  & 47 & 1.0 & 752 $\pm$ 31 \\
E2 & 2010-04-03 &  4.6   & -25.1& 77  & 50 & 1.2 & 824 $\pm$ 29 \\
E3 & 2011-04-17 & -177.1 & 8.9  & 72  & 52 & 1.1 & 1312 $\pm$ 51 \\
E4 & 2011-06-21 & -20.8  & 9.5  & 70  & 70 & 1.5 & 1006 $\pm$ 38 \\
E5 & 2011-08-03 & 15.4   & 16.4 & 68  & 41 & 0.8 & 1294 $\pm$ 56 \\
E6 & 2012-02-23 & 67.2   & 24.4 & 139 & 82 & 3.5 & 445 $\pm$ 12  \\
\hline
\end{tabular}
\tablenotetext{}{Note: $\phi$ and $\theta$ are the longitude and latitude of the
3D CME propagation direction, respectively. Positive/negative values of $\phi$
mean westward/eastward propagation with respect to the Earth.
$W_f=2(\alpha+\delta)$ and $W_e=2\delta$ are the face-on and edge-on angular
width, respectively. $\Omega=W_fW_e=4\delta(\alpha+\delta)$ approximates the
solid angle of a CME. $v_{\mathrm{avg}}$ is the
average speed of the CME from the linear fit to the observed height-time profile.}
\label{tb_1}
\end{table}

\begin{table}
\caption{Derived parameters of CME mass at different heights, solar wind
parameters and its mass contribution}
\centering
\begin{tabular}{ccccccccccc}
\hline
No. & $m_0$ & $h_{occ}$ & $m_1$ & $h_1$ & $m_2$ & $h_2$ & $n_{sw}$ & $v_{sw}$ & $m_{s}$ & $f_s$ \\
    &$\log_{10}$(g)&(\(\Rs\)) &$\log_{10}$(g)&(\(\Rs\))&$\log_{10}$(g) &(\(\Rs\))& (cm$^{-3}$) & (km s$^{-1}$)& $\log_{10}$(g)& \\
E1  &15.17 & 2.83 & 15.27 & 7.74 & 15.37 & 14.2 & 869 & 263 & 15.09 & $< 2.55$ \\
E2  &15.38 & 2.34 & 15.48 & 7.76 & 15.60 & 14.3 & 657 & 376 & 15.01 & $< 1.06$ \\
E3  &15.35 & 1.82 & 15.63 & 7.01 & 15.86 & 15.7 & 824 & 243 & 15.32 & $< 0.70$ \\
E4  &15.45 & 2.69 & 15.62 & 7.97 & 15.78 & 15.8 & 696 & 355 & 15.20 & $< 0.85$ \\
E5  &15.42 & 2.62 & 15.68 & 8.46 & 15.85 & 15.6 & 878 & 226 & 15.15 & $< 0.62$ \\
E6  &15.65 & 2.79 & 15.83 & 7.51 & 16.01 & 15.3 & 610 & 338 & 15.28 & $< 0.55$ \\
\hline
\tablenotetext{}{Note: $m_0$ is the modeled initial apparent mass of a CME below the
effective occulter height $h_{occ}$. $m_1$ is the apparent mass at the
height $h_1$ where the occulting effect begins to be negligible. $m_2$ is the
final apparent mass before a CME leaves COR FOV, and $h_2$ is the corresponding
height. $n_{sw}$ and $v_{sw}$ are the ambient solar wind density and speed at
$h=20 $\(\Rs\). $m_s$ is the solar wind mass piled up around the CME from $h_1$
to $h_2$. $f_{s}=\frac{m_s}{m_2-m_1}$ measures the contribution of the solar wind pileup
to the apparent mass increase in the height range from $h_1$ to $h_2$. }
\end{tabular}
\label{tb_2}
\end{table}

\end{document}